\documentstyle[aps,pre,twocolumn,epsf]{revtex}
\begin{document}
\def\be {\begin{equation}}
\def\ee {\end{equation}}
\def\bee {\begin{eqnarray}}
\def\eee {\end{eqnarray}}
\def\N {{\cal N}}
\def\z {\zeta}
\def\zk {\zeta_k}
\def\OP {\tensor P}
\def\B.#1{{\bbox{#1}}}
\renewcommand{\thesection}{\arabic{section}}
\title{{\rm J.Stat.Phys, submitted.   \hfill  Version of \today}\\~~
\\Dynamics of Finger
Formation in Laplacian Growth without Surface Tension}
\author {Mitchell J. Feigenbaum$^{*}$,
Itamar Procaccia$^{**}$ and Benny Davidovich$^{**}$ }
\address{$^*$The Rockefeller University, 1230 York Ave.New York, NY
10021,\\$^{**}$Department of~~Chemical Physics, The
Weizmann  Institute of Science, Rehovot 76100, Israel.}
\maketitle
\begin{abstract}
We study the dynamics of ``finger"
formation in Laplacian growth without surface tension in a channel
geometry (the Saffman-Taylor problem). Carefully determining
the role of boundary geometry (resulting in reflection symmetry, not
periodicity),
we construct field equations of motion, these central to the analytic power
we can
here exercise. Initially we consider an explicit analytic class of maps to
the physical space, a basis of solutions for infinite fluid in an infinitely
long channel, characterized by meromorphic derivatives. We dynamically
verify that these maps
never lose analyticity in the course of temporal evolution, thus justifying
the
underlying machinery. However, the great bulk of these solutions can lose
conformality
in time, this the circumstance of finite-time singularities.
By considerations of the nature of the analyticity of all these solutions,
we do, however,
show that those free of such singularities inevitably result in a {\em
single} asymptotic
``finger". This is purely nonlinear behavior: the very early ``finger"
actually already
has a waist, this having signalled the end of any linear
regime. The single ``finger" has nevertheless an arbitrary width
determined by initial conditions. This is in contradiction
with the experimental results that indicate selection of a finger
of width 1/2. In the last part of this paper we motivate that
such a solution can be determined by the boundary conditions when the
fluid is finite. This is a strong signal that
{\em finiteness} is determinative of pattern selection.
\end{abstract}
\pacs{PACS numbers
64.60Ak,05.20Dd,41.10.Dq}
\section{Introduction}
The aim of this paper
is to offer some careful analysis of the initial value problem
associated with Laplacian growth without surface tension in an infinite
channel geometry. It is well known for quite some time that this
problem possesses {\em asymptotic} solutions in the form of a ``finger"
whose width $\lambda$ may be taken apparently arbitrarily \cite{58ST}. In
experiments which are performed in a finite channel and in which
the effective surface tension is very small a single width ($\lambda=1/2$) is
selected\cite{58ST,87TZL}. Previous work explained how a unique steady
solution is selected by the existence of surface
tension \cite{86Shr,86HL,86CDHPP,87Tan}. That work never explored the question
how this solution arises from arbitrary initial conditions.

Here we consider the
initial value problem, and explain how a very wide class of
initial conditions ends up with an asymptotic solution of an interface in the
form of a single
finger. Our initial conditions may include very complicated interfaces,
indeed having many features. It is important to understand the generic
dynamical mechanism that is responsible for the generation of a single
finger that becomes an asymptotic solution. We show that in the infinite
channel the width of the finger is not selected (notwithstanding some
recent claims to the contrary \cite{98Min}).

We next remark that the experiments in which selection is observed
include, in addition to surface tension, also a second boundary condition.
One can set up the experiment in various ways, like ending
the material fluid at
atmospheric pressure, or connect it to a pump that sustains the flow. We
discuss the various boundary conditions that can be imposed on the
second boundary, and show that some natural choices impose a unique
solution with width 1/2. This discussion serves as an introduction to
a substantial treatment of finiteness in a forthcoming paper \cite{99Fei}

The structure of the paper is as follows: in Sect.2 we recall the
analytic theory of Laplacian growth in channel geometry. The novel
aspect of our treatment is the introduction of the equation of motion
for the conformal map that explicitly takes into account reflection
symmetry of the solutions in the complex plane. The significant virtue of
our method is that
it totally obviates any need for Hilbert transforms and so forth that are
required in the usual, pure boundary treatment.
.In Sect.3 we show that analytic initial data remain analytic for all
available time. On the
contrary, conformality can be lost in finite time, and in Sect.5 we discuss
how this may
happen in generic situations, and discuss how a regulation, such as surface
tension, accomplishes the cure of this disease. In Sect.4 we come to understand
that any generic initial condition has a single Saffman-Taylor finger
asymptotics. In Sect.6
we discuss finite boundary conditions. We observe that the imposition of a
downstream
boundary  condition may uniquely select a solution of width 1/2.
In Sect.7 we offer a summary and a discussion.

\section{Analytic solutions of Laplacian Growth in Channel Geometry}
\subsection{The physical problem and the mathematical formulation}
We are interested in
the so-called Saffman-Taylor problem of determining the motion of an
interface $\gamma(t)$ between two fluids of different viscosities in a
Hele-Shaw cell. We can think of air displacing oil as a standard example
\cite{88Pel}. The cell is a channel made of two long rectangular plates
displaced by a small distance $b$. We chose $x$ to denote the long lateral
coordinate, whereas $y$ denotes the transversal direction of
the cell, $0\le y\le\pi$ in suitable units, (see Fig.1a). When the gap $b$
is considerably smaller than the lateral width of the cell, and non-slip
boundary conditions are taken at the upper and lower plates, then the
velocity field  $\B.v$ in the driven fluid satisfies Darcy's
law
\begin{equation}
\B.v={-b^2\over 12\mu}\B.\nabla p \ ,
\label{darcy}
\end{equation}
where $p$ is the pressure field and $\mu$
the viscosity. Because of the assumed very small viscosity of the
driving fluid, its pressure is almost constant (taken to be zero),
while in the driven fluid, by virtue of incompressibility, $\B.\nabla
\cdot\B.v=0$, the pressure is harmonic:
\begin{equation}\Delta p=0 \
. \label{laplace}
\end{equation}
The boundary conditions on the
interface are determined by first requiring the non-penetrability of
the two fluids in contact. This means an equality of the normal
velocity of the interface $\B.v_i\cdot\B.n$ and of the normal velocity
of the fluid at the interface. Secondly,
the pressure of the fluid at the interface is given by the relation
\begin{equation}
p_{air}-p=\sigma \kappa \ , \label{dyncon}
\end{equation}
where $\sigma$ and $\kappa$ are interfacial
surface tension and curvature respectively. On the lateral
walls ($y=0,\pi$) the normal velocity of the fluid vanishes, and as
$x\to \infty$, far ahead of the interface, the flow
is taken as uniform, parallel
to the $x$ axis, and of magnitude $V$. Here we take $\sigma=
0$. The boundary condition (\ref{dyncon}) now simplifies to
taking constant pressure on the interface:
\begin{equation}
p=0  ~{\rm at~ the~ interface~\gamma(t)}; ~\sigma=0 \
. \label{peq0}
\end{equation}
In consequence of Eq.(\ref{darcy}) we also need to
require that the normal derivative of the pressure vanishes at the
lateral walls. This means that the interface must be normal to the two
lateral walls. As the flow is approximately two-dimensional and obeys
Laplace's equation, in the usual fashion one produces an analytic
function $\tilde h'(z)$,
\begin{equation}
\tilde h'(z)=-\partial_x p+i
\partial_y p=v_x-iv_y\equiv \bar \B.v \
,\label{h'z}
\end{equation}
where $z=x+iy$. (With $\Delta p=0$, $\tilde
h'(z)$ satisfies the Cauchy-Riemann conditions.) The integral of $\tilde
h'(z)$ is
\begin{equation}
\tilde h(z)=-p(z)+is(z) \ ,
\label{tildh}
\end{equation}
with $s(z)$ the harmonic function conjugate
to $-p(z)$. The function $s(z)$ is typically multivalued. Consider
now an arbitrary
curve $\gamma$, and denote by $\Phi_\gamma(t)$ the time dependent flux
that crosses $\gamma$. It is convenient to consider $V_\gamma(t)$,
the mean channel velocity, by dividing $\Phi$ by the constant channel
width $\pi$. With $\hat n$ denoting the right-handed
normal
\begin{equation}
V_\gamma(t)=\frac{1}{\pi}\int_\gamma v_n d\ell \ , \label{intcond}
\end{equation}
where $d\ell$ is an arc-length differential. Denoting
$\hat \B.u$ the unit vector orthogonal to the plane we can write
\bee
v_n d\ell&=&\B.v\cdot(d\B.\ell\times\hat\B.u)=\B.v\times \B.d\B.\ell\cdot \hat
\B.u\nonumber\\&=&v_xdy-v_ydx ={\rm Im}\bar vdz \ .
\eee
We observe that
\begin{equation}
\frac{1}{\pi i}\bar vdz=-\frac{i}{\pi}\B.v\cdot\B.d\B.\ell
+\frac{1}{\pi}v_nd\ell=\frac{i}{\pi}dp+\frac{1}{\pi}
v_n d\ell \ ,
\end{equation}
and
\begin{equation}V_\gamma(t)=\frac{1}{\pi
i}\int_\gamma\bar v dz-\frac{i}{\pi}\delta_\gamma p \ ,
\end{equation}
where $\delta_\gamma p$ represents the pressure {\em difference}
between the end points of $\gamma$. In particular, the flux
crossing a curve of constant pressure $p=p_0$ satisfies
\begin{equation}
V_{p_0}(t)=\frac{1}{\pi i}\int_{p_0}\bar v dz\ . \label{Vconstp}
\end{equation}
We will later need to consider sinks  at finite distances.
In preparation for this, consider two
curves, $\gamma_1$ and $\gamma_2$, each connecting the
two boundaries, and observe that
\begin{equation}
V_{\gamma_2}-V_{\gamma_1}=\frac{1}{\pi i}\oint\bar v dz \ ,
\end{equation}
where the closed curve is the boundary of the domain
$\Omega$ which is delineated by $\gamma_1$, $\gamma_2$ and the two walls.
If $\bar v$ is analytic in $\Omega$,
$V_{\gamma_1}=V_{\gamma_2}$. On the other hand if there are sinks in
$\Omega$, we can write
\begin{eqnarray}
\bar v=\bar v_{\rm ana}+\sum{a_i/2\over z-z_i}\nonumber\\
V_{\gamma_1}-V_{\gamma_2}=\sum a_i \ ,
\end{eqnarray}
and each such pole sinks $a_i$ worth of the flux
crossing $\gamma_1$ into $\Omega$. Returning to the flux across a
constant pressure line we note from Eqs.(\ref{Vconstp}) and
(\ref{tildh}) that $V_{p_0}=\delta_{p_0}s/\pi$, or
\begin{equation}
\delta_{p_0}{s\over V_{p_0}(t)}  = \pi  \ . \label{dels}
\end{equation}
Since we are assuming a flux of fluid
incident from the left, it is evident that there must be a sink of
equivalent strength either at infinity or at some finite location. In
particular the pressure approaches $-\infty$ at the sink. Provided
there is just one sink, then for each value of $p$ there will generally
be just one corresponding physical curve. Then, at each instant of
time, the flux crossing every curve of constant pressure is
identical. Defining a complex variable
\begin{equation}
\zeta \equiv {-p+is\over V(t)} \ ,
\end{equation}
where $V(t)$ is the common value of
$V_{p_0}$ for all constant pressure lines, we now see by
Eq.(\ref{dels}) that $\tilde h/V(t)$ maps the physical channel into an
identical region of $\zeta$ space. That is, the map
\begin{equation}
\zeta=h(z,t)\equiv\tilde h(z,t)/V(t)
\end{equation}
can conformally map the strip to itself. In particular
\begin{equation}
h'={\bar v\over V(t)} \ , \label{h'}
\end{equation}
replaces Eq.(\ref{h'z}). Laplace's equation
for the pressure is written as $\Delta {\rm Re}\zeta =0$, and is
automatically still obeyed since $V$ is purely a function of
time. However, we must insist now that the interface is $p=0$ and not
any other constant, not to have Re$\zeta$ time dependent on the
interface. (The {\em physical} pressure can be any $p_0(t)$.
Since only differences of pressure matter, we simply subtract $p_0(t)$
to normalize to $p=0$ at the interface.)

\begin{figure}
\epsfxsize=8.0truecm
\hskip 0.5cm
\epsfbox{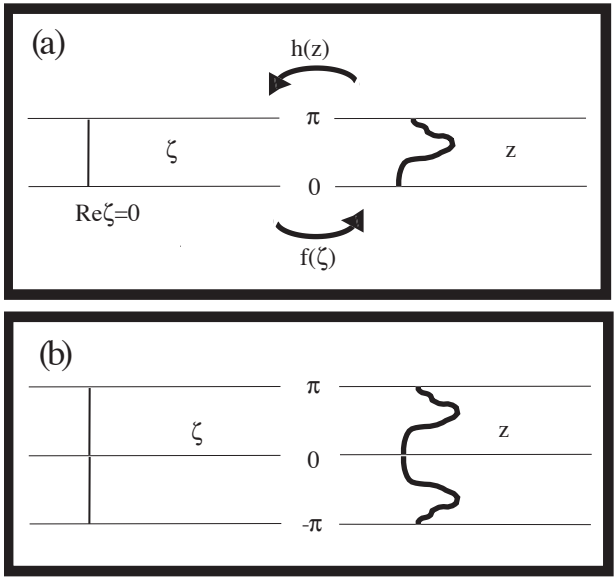}
\vskip 0.5cm
\caption{(a) The physical channel in $z$ space with the interface
satisfying no-flux
on the lateral walls is mapped onto the mathematical channel
in $\zeta=h(z)=-p(z)+is(z)$. (b) Using the reflection
symmetry the problem is ``doubled", the interface in the doubled channel
satisfies periodic boundary conditions, and the expected physical
solutions is {\em two fingers}.}
\label{channel}
\end{figure}

The relations between the physical channel, the mathematical
strip and $h(z)$ are summarized graphically in Fig.(\ref{channel})a.

\narrowtext

\subsection{The analytic map $f$ and its equation of motion}
Having defined the map $h$
we notice that it is inconvenient that the boundary of its domain
$\gamma(t)$ is at the moment unknown and potentially
complicated. However its image in $\zeta$ space is
elementary: Re$\zeta=0$. Accordingly it is natural to invert the
discussion and consider a map $f$ from $\zeta$ to $z$. Assuming $h$ is
{\em conformal} ($h'\ne 0$) on the physical domain), $h^{-1}$ exists,
which is precisely the desired analytic (in the physical domain) $f$
($f\equiv h^{-1}$). From this point onwards when we say ``analyticity",
we mean the analyticity of $f$ (equivalent to the conformality of
$h$). As well, when we say ``conformality" we mean the conformality of
$f$ (equivalent to the analyticity of $h$). Should $f$'s analyticity
fail, then the setup is inapplicable; should $f$'s {\em conformality}
fail, then $\Delta p=0$ no longer holds, and the dynamics has created
singularities.  We will show in Sec.3 that $f$ remains analytic for all
times. The serious issue of $f$'s conformality is beyond our full
understanding, but we shall illuminate the issue. $f(\zeta,t)$ describes
a flow with boundary conditions $p=0$ provided that the
interface $\gamma(t)$ develops to the interface $\gamma(t')$ for any
later time $t'$ under transport by $\B.v=\bar h'$. This requirement
leads to the equations of motion \cite{45Pol,45Gal,84SB,94DMW}. Consider
any point $\zeta_0\in \Omega$. $\zeta_0$ serves as a Lagrangian label
for the fluid point at $f(\zeta_0,0)$. However, $f(\zeta_0,t)$ is {\em
not} this same fluid point at time $t$. Define
\begin{equation}
z=Z(\zeta_0,t)
\end{equation}

as the moving point labeled by $\zeta_0$, with $Z(\zeta_0,0)=f(\zeta_0,0)$.
There exist at each time $t$ a map $\zeta_0\to \zeta(\zeta_0,t)$ such that
$Z(\zeta_0,t)=f(\zeta(\zeta_0,t),t)$. By definition
\begin{eqnarray}
Z_t(\zeta_0,t)&=&f'\zeta_t+f_t=v_x+iv_y\nonumber\\&=&V(t)
\overline{h'(f(\zeta,t)}={V(t)\over
\overline{f'(\zeta,t)}} \ .\label{zomb}
\end{eqnarray}
Accordingly $\zeta$ satisfies the differential equation
\begin{equation}
|f'|^2\zeta_t+f_t\bar{f'}\equiv V(t) \
. \label{eqzeta}
\end{equation}
We observe now that the boundary
$\gamma(t)$ flows into itself. Thus for each $\zeta_0$ on the boundary
(Re$\zeta=0$ at $t=0$), $\zeta(\zeta_0,t)$ is also on the boundary, and
so Re$\zeta(\zeta_0,t)\equiv 0$ for Re$\zeta_0=0$. But then Re$\zeta_t=0$ and
taking the real part of (\ref{eqzeta}) we obtain the usual result
\cite{45Pol,45Gal,94DMW}
\begin{equation}
{\rm Re}(f_t\bar{f'})=V(t) \ ,
\quad {\rm on}~{\rm Re}\zeta=0 \ . \label{lge}
\end{equation}
In previous work $V(t)$ is taken identically equal to
1. Eq.(\ref{lge}) generalizes the standard result to the case of
variable flux which is necessary with finite boundary conditions.

\subsection{Exponentiation and reflection}
\begin{figure}
\epsfxsize=9.0truecm
\epsfbox{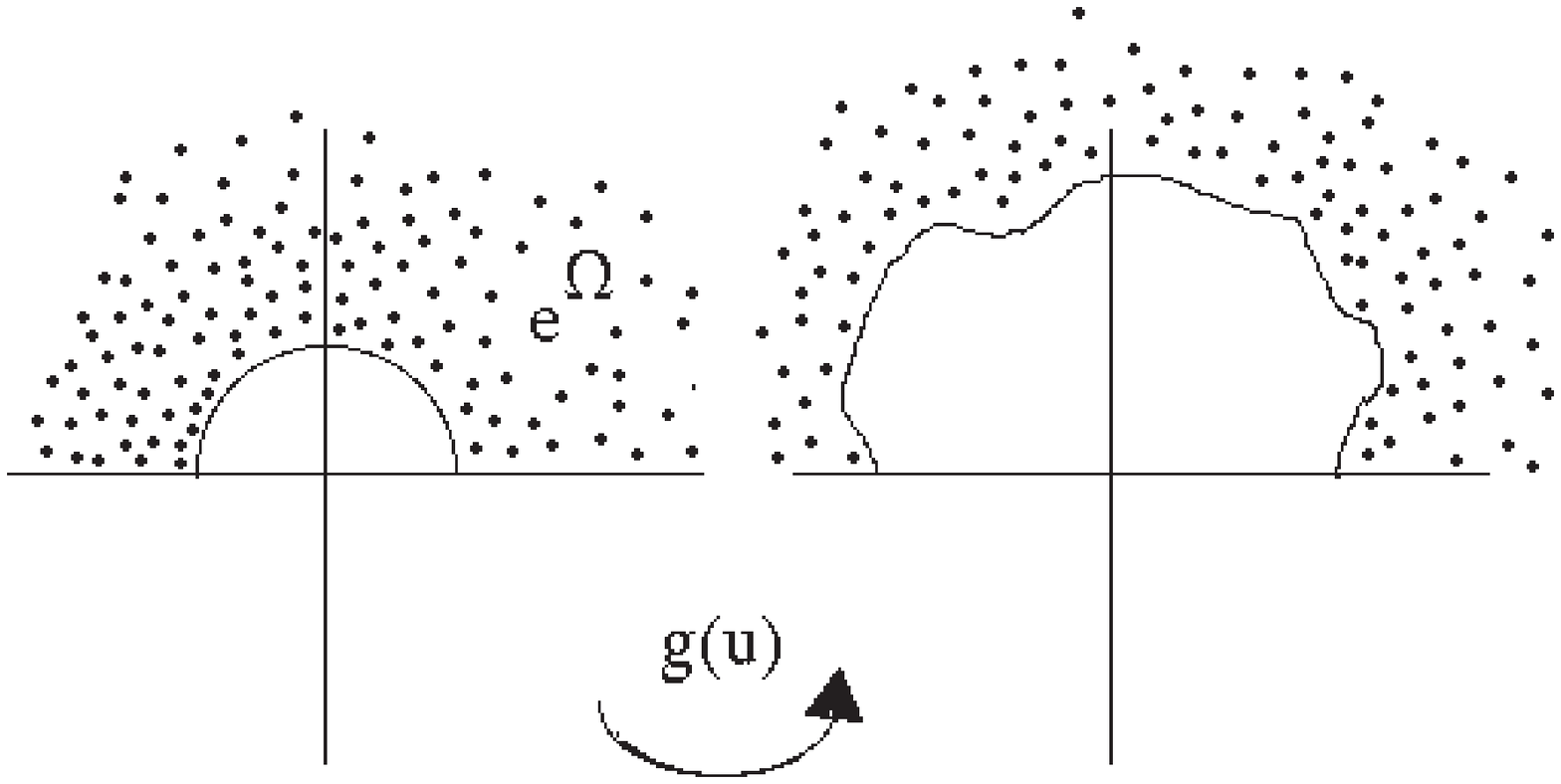}
\caption{The exponentiated domain and its mapping $g(u)$ to the exponentiated
physical strip}
\label{exponent}
\end{figure}

Geometrically, the strip $\Omega$, Re$\zeta\ge 0$, $0<{\rm
Im}\zeta<\pi$, is less felicitous then a circular domain, with
``infinity" (Re$\zeta\to \infty$) {\em the} point at infinity. It is
natural to write
\begin{eqnarray}
u&\equiv& e^\zeta\ , w\equiv e^z \ ; \nonumber\\
w&\equiv& g(u)\ , \quad f(\zeta)=\ln g(e^\zeta)\ ,
\label{uw}
\end{eqnarray}
so that $ e^\Omega$ is the entire upper
half-plane minus the unit disc ($|u|\ge 1$, Im$u>0$), see the shaded
region in Fig.(\ref{exponent}). Under $f$'s exponential conjugate, $g$,
$e^\Omega$
maps to Im$w>0$:

\narrowtext

$g$'s boundary value on $|{\rm Re}u|\ge 1$, Im$u\to 0^+$, is {\em real},
and just the exponential of pressure along the walls. With no
shocks, $p(x,0)$ is continuous. So $g$ takes real rays continuously to
real rays. As such it immediately defines its analytic
continuation $g(u)\equiv \overline{g(\bar u)}$ for Im$u<0$, and this
reflection symmetric continuation is analytic in the {\em entire}
$u$-plane for $|u|>1$. With ln and exp both also
reflection-symmetric, so too must be $f$:
\begin{equation}
f(\zeta)=\overline{f(\bar\zeta)}\ ,
\label{barbar}
\end{equation}
as well as $f'$ and $f_t$. We shall have
numerous occasions to utilize this symmetry. Note that the continued $f$
has a domain equivalent to a doubling of the physical problem such that
the interface and fluid now enjoy analytically periodic boundary
conditions from $-\pi$ to $\pi$. Of course, the physical problem remains
in the half strip, but this extra symmetry dictates profound
restrictions on $f$'s analyticity structure \cite{93Tan}.
\subsection{Reflection symmetric equations of motion}
Let us now utilize reflection symmetry
to transform Eq.(\ref{lge}) into a field equation valid throughout the
body of the fluid. Using (\ref{barbar}) we can rewrite Eq.(\ref{lge})
in the form
\begin{equation}
2V(t)=f_t(\zeta)f'(\bar\zeta)+f_t(\bar\zeta)f'(\zeta) \ ,
\end{equation}
and noticing that Re$\zeta=0$ implies that $\bar\zeta=-\zeta$ we
have
\begin{equation}
f_t(\zeta)f'(-\zeta)+f_t(-\zeta)f'(\zeta)=2V(t) \
. \label{lgesym}
\end{equation}
It is important to note that this
equation holds not just for Re$\zeta=0$, but for all $\zeta$ for which
f is analytic. The reason is as follows: solve for $f(\zeta,t)$ from
Eq.(\ref{lge}). Compare to the solution of Eq.(\ref{lgesym}), which we
denote temporarily as $\tilde f(\zeta,t)$. The two functions
$f(\zeta,t)$ and $\tilde f(\zeta,t)$ are the same on the boundary
Re$\zeta=0$, and since they are analytic they are the same
functions. We can thus use (\ref{lgesym}) instead of (\ref{lge}) for
any $\zeta$. Eq.(\ref{lgesym}) relates $f(\zeta)$ to $f(-\zeta)$ which
will prove to be of significant value for the forthcoming analysis. In
particular, when $f$ has the elementary analytic behavior $f(\zeta)\sim
\zeta$ as Re$\zeta\to\infty$, it follows from (\ref{lgesym}) that $f$
is analytic as Re$\zeta\to -\infty$. (This is the usual case for the
consideration of a sink at $\infty$.)
\subsection{A class of solutions}
There exist, sprinkled throughout the literature \cite{84SB,94DMW,86How},
considerations of a class of
solutions to (\ref{lgesym}) with a sink at $\infty$, with moving
singularities, most naturally written in $u$-space as
\begin{equation}
g(u)=e^{\beta(t)}u~\prod_{k} \left(1-{a_k(t)\over
u}\right )^{\alpha_k} \ .\label{g(u)}
\end{equation}
It is easy to see
that for such $g$ the corresponding $f_t$ and $f'$ simply have poles in
$e^\zeta$. From the form of (\ref{lgesym}) we then see that the class
of functions $g$ exhausts the solutions of (\ref{lgesym}) within the
class of rational functions of $e^\zeta$. We will verify (as has been
commented upon in some of the literature \cite{94DMW}) that the
equations of motion are satisfied with $\alpha_k$ constant in
time. Contrary to some of the literature we consider the $\alpha_k$
arbitrary complex numbers rather than just real integers
\cite{84SB}. Further, paying attention to the analyticity structure
imposed by reflection symmetry, it follows that $\beta(t)$ is real,
$|a_k(t)|<1$, and for each complex $\{a_k(t),\alpha_k\}$ pair, there
must also be a corresponding pair with both conjugated. Those $g$'s
with just one $a$ (which is real and constant in time) comprise the
Saffman-Taylor
solutions. Numerical explorations in the literature appear to indicate
\cite{94DMW} that richer $g$'s replicate a large class of interface
motions. (From a numerical point of view a rich enough collection of
$a$'s surely serves as a basis.) Some of our comments will follow
directly from the equations of motion, whereas others pertain only to
arbitrary solutions of the form (\ref{g(u)}). By the conjugacy
(\ref{uw}) we now have for $f$
\begin{equation}f(\zeta,t)=\beta(t)+\zeta+
\sum_{k=1}^N\alpha_k\ln\left(1-e^{\zeta_k(t)-\zeta}\right)\ ,
\label{exact}
\end{equation}
where $a_k(t) \equiv \exp(\zeta_k(t))$,
Re$\zeta_k(t)<0$. We shall suspend until Sect.3 the equations of motion
implied by (\ref{lgesym}) for the parameters of $f$, and further comments
on the origin of class (\ref{exact}) in subsection G to follow.
\subsection{Elementary flow solutions}

Our equations of motion allow for the ready production of a
variety of solutions. Consider first the class of solutions with a fixed
shape interface uniformly translating in time, so that we can take
$V(t)\equiv 1$. The interface is
\begin{equation}
z_{\rm int}(s,t)=f(is,t)=\beta(t)+F(is) \ . \label{zint}
\end{equation}
Using (\ref{lgesym}),
\begin{eqnarray}
f&=&\beta(t)+F(\zeta);\quad f_t=\dot \beta(t),\quad
f'=F'(\zeta)\nonumber\\
\frac{2}{\dot\beta}&=&F'(\zeta)+F'(-\zeta) \
. \label{2betadot}
\end{eqnarray}
But then, $1/\dot\beta\equiv \lambda$,
a constant, and $\beta=t/\lambda$. From (\ref{2betadot}) we deduce
\begin{equation}
F(\zeta)-F(-\zeta)=2\lambda\zeta \ ,
\label{F-F}
\end{equation}
or $f(\zeta,t)-f(-\zeta,t)=2\lambda\zeta$. Setting $\zeta=is$, we
find
\begin{equation}
f(is,t)-f(-is,t)=2i\lambda s \ .
\end{equation}
By reflection  symmetry $f(-is)=\bar f(is)$, i.e. Im$f(is,t)=\lambda s$,
or
\begin{equation}y(s,t)=\lambda s \ .
\end{equation}
This solution represents an interface that occupies a constant channel
fraction, $\lambda$, sensible for $\lambda \in (0,1]$. This also
implies that the interface is a graph of $(x(y),y)$, and so conformal
for those $\lambda$'s since $|f'|>|y'|=\lambda$. Writing
$F=\lambda\zeta+E(\zeta)$, Eq.(\ref{F-F}) is
\begin{equation}E(-\zeta)=E(\zeta) \ .
\end{equation}
The general solution is
\begin{equation}
f=\frac{t}{\lambda}+\lambda\zeta+E(\zeta) \ ,
\end{equation}
with $E$ an even function of $\zeta$ compatible with
full channel width:
\begin{equation}
\lambda+\frac{1}{\pi i}\left(E(-p+i\pi)-E(-p)\right)=1
\ ,\end{equation}
for those curves of constant pressure within the
physical fluid ($p<0$) which span the free channel. Since $E$ is even,
if $E$ has a singularity at $\zeta_k$ with Re$\zeta_k<0$, then it must
also possess an identical singularity at Re$\zeta_k>0$, and so $f$ is
no longer analytic in the entire physical fluid, implying the existence
of stagnation points or infinite velocities within the fluid itself. Within
the class of
solutions $g$,
\begin{eqnarray}
&&f=\frac{t}{\lambda}+\lambda\zeta+\sum
\alpha_k\ln(1-e^{\zeta_k-\zeta})(1-e^{\zeta_k+\zeta})\nonumber\\
&&{\rm Re}\zeta_k\ge 0~({\rm by ~definition})\
. \label{bydef}
\end{eqnarray}

It is expedient, to facilitate the generation of elementary solutions,
to consider those solutions which are ``periodic" over the {\em physical}
channel:
\begin{equation}
f(\zeta+i\pi)=f(\zeta)+i\pi \ .
\end{equation}
Defining now double-width variables
\begin{equation}
\tilde \zeta = 2\zeta -i\pi;\quad \tilde z=2z-i\pi
\end{equation}
so that the physical channel $0\le$Im$z\le\pi$ is mapped to
$-\pi\le$Im$\tilde z\le \pi$, we consider
\begin{eqnarray}
\tilde z&=&2f\left(\frac{\tilde\zeta+i\pi}{2}\right)-i\pi\equiv
\tilde f(\tilde \zeta)\nonumber\\
f(\zeta)&=&\frac{1}{2}\tilde f(2\zeta-i\pi)+\frac{i\pi}{2} \ .
\end{eqnarray}
$\tilde f$ now has $2\pi i$ periodicity, and is again reflection symmetric
since $f$ is:
\begin{eqnarray}
&&\bar{\tilde f}(\bar{\tilde \zeta})= 2 \bar f \left
(\frac{\bar{\tilde \zeta}+i\pi}{2}\right)+i\pi=
2f\left(\frac{\tilde \zeta-i\pi}{2}\right)+i\pi\\
&&=2f\left(\frac{\tilde \zeta+i\pi}{2}-i\pi\right)+i\pi=
2f\left(\frac{\tilde\zeta+i\pi}{2}\right)-i\pi=\tilde f(\tilde \zeta)\ .
\nonumber
\end{eqnarray}
Conversely, any $2\pi i$ periodic reflection symmetric $\tilde f$ determines
a $\pi i$ periodic reflection-symmetric $f(\zeta)$. (Such an $f$ is too
symmetric in that the upper and lower channel walls have perfectly synchronized
flows along them, a condition not enforced by the physical data.) The virtue
of this device is that an $\tilde f$ with just one symmetric pair of\
singularities produces an $f$ with {\em two} symmetric pairs of
singularities, and so embraces more complicated flows that does just
one symmetric pair.

So far as time-dependence is concerned it is natural to take
\begin{equation}
f(\zeta,t)=\frac{1}{2}\tilde f(2\zeta -i\pi,2t)+i\frac{\pi}{2}
\quad (\tilde t\equiv 2t) \ ,
\end{equation}
so that
\begin{equation}
f'(\zeta,t)=\tilde f'(\tilde \zeta,\tilde t)\ , \quad f_t(\zeta,t)=\tilde f
_{\tilde t}(\tilde \zeta,\tilde t) \ .
\end{equation}
Also, by $2\pi i$ periodicity of $\tilde f$, $f'(-\zeta,t)=\tilde f'(-\tilde
\zeta,\tilde t)$ and similarly for $f_t$. With $\tilde V(\tilde t)\equiv
V(t)$, the equations of motion are covariant. So, all we need do is forget the
tildes, solve the simpler problem, and then conjugate it back to the
physical space. In the sequel we neglect the last step as an ellipsis
the reader readily can fill in. There are reasons to be wary of the extra
symmetry in the solutions, and we shall point out this fact when it arises.

The simplest possibility of (\ref{bydef})
is one $\zeta_k=\pi i$:
\begin{eqnarray}
f&=&\frac{t}{\lambda}+\lambda\zeta+
\alpha\ln(1+e^{-\zeta})(1+e^{\zeta})\nonumber\\&=&\frac{t}
{\lambda}+(\lambda+\alpha)\zeta+2\alpha\ln(1+e^{-\zeta}) \ .
\end{eqnarray}
For full channel width for Re$\zeta>0$, we require
$\lambda+\alpha=1$, or
\begin{equation}
f=\frac{t}{\lambda}+\zeta+2(1-\lambda)\ln(1+e^{-\zeta})
\quad(f'(-\infty)=2\lambda-1) \ . \label{10}
\end{equation}
These are precisely the Saffman-Taylor
solutions for a finger of width $\lambda$, and the only such solutions
of our class analytic for all Re$\zeta>0$.

\begin{figure}
\epsfxsize=9.0truecm
\epsfbox{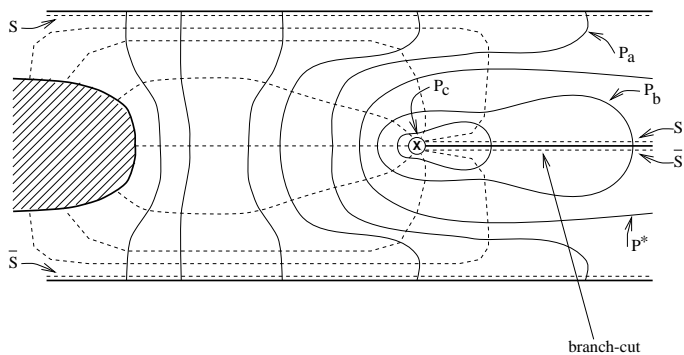}
\caption
{Schematic stream and pressures lines in a channel with an advancing finger
(shaded
region) and a sink (denoted by $\times$). The stream and pressure lines are
indicated by dashed
and continuous curves respectively. The pressure line with $p=p^*$
separates pressure
lines that connect the walls (like $p_a$) and those which form closed
curves around
the sink (like $p_b$). The stream lines $s=\pi, \bar s=-\pi$ graze the walls
$y=\pm \pi$ respectively from $x=-\infty$ to $x=\infty$ where they curve around
and return along $y=0^+$ and $y=0^-$ respectively from $x=\infty$ to the sink,
forming a branch cut.}
\label{mitchsink}
\end{figure}

The next simplest solution is
\begin{eqnarray}
f&=&\frac{t}{\lambda}+\zeta+2(1-\lambda)
\ln(1+e^{-\zeta})\nonumber\\&-&\alpha\ln(1+e^{-\zeta_*-\zeta})
(1+e^{-\zeta_*+\zeta})\quad(\zeta_*>0) \ . \label{nss}
\end{eqnarray}
Notice that for $-\zeta_*<{\rm
Re}\zeta<\zeta_*$, the channel is fully occupied with flux, but for
Re$\zeta>\zeta_*$ the entire asymptotic flux occupies only the fraction
$1-\alpha$ of the whole channel. The most interesting such case is
$\alpha=1$, when all flux is sunk in the point $f(+\infty)=z_{\rm
sink}$:
\begin{equation}
z_{\rm sink}(t)=\frac{t}{\lambda}\ , \quad
f'(+\infty)=0 \ . \label{zsink}
\end{equation}
The constant pressure contours for $0>p>-\zeta_*$ span the channel, but
those for $p<-\zeta_*$ are closed curves surrounding the sink at $z_{\rm
sink}$. The separatrix between them at $\zeta=\zeta_*+is$ has the form
\begin{eqnarray}
z_*&=&\frac{t}{\lambda}+\zeta_*+is+2(1-\lambda)\ln(1+e^{-\zeta_*-is})
\nonumber\\&-&\ln(1+e^{-2\zeta_*-is})-\ln(1+e^{is})\ .
\end{eqnarray}
For $\zeta_*\gg 1$
\begin{equation}z_*\approx \frac{t}{\lambda} +\zeta_*+is-\ln(1+e^{is}) \ ,
\end{equation}
and is an upstream-pointing Saffman-Taylor finger of width $2(1-\lambda)=1$,
or $\lambda=1/2$, by comparison to (\ref{10}), see Fig.\ref{mitchsink}.
This is
amusing, and suggests that with an enforced symmetry between $\zeta$
and $-\zeta$, then (\ref{zsink}) would be $f'(-\infty)=0$, so that if
(\ref{10}) were asymptotically valid, then a $1/2$ width finger for the
interface itself could be implied. We will return to these matters later.

Let us now consider some elementary solutions with variable flux
$V(t)$. The simplest solution arises when the pressure profiles are a
function of $x$ but not of $y$, $p=p(x,t)$. This implies, by
analyticity, that
\begin{equation}
f=\beta(t)+\zeta \ , \label{bet+zet}
\end{equation}
where the coefficient 1 of $\zeta$ specifies a channel fully occupied
with flux. We now pose {\em finite} boundary conditions, namely that
$p=p_a$ (atmospheric pressure) on $x=L$. We denote $p_g=|p_a|$, and
write $2L=f(\zeta)+f(\bar\zeta)$, on
$2p_g/V=\zeta+\bar\zeta$. Accordingly
\begin{equation}
2L=f(\zeta)+f(\frac{2p_g}{V}-\zeta)=2\beta+\frac{2p_g}{V}\ , \label{2L=}
\end{equation}
or $L=\beta+p_g/V$. Using the fact that in this case
$\dot \beta=V$, we need to solve the differential equation
\begin{equation}
\beta\dot\beta-L\dot\beta+p_g \ . \label{betbetdot}
\end{equation}
The solution is $\beta^2-2L\beta+2p_gt=const$. Choosing initial
conditions such that the interface is at zero when $t=0$ determines
$x_{\rm int}=\beta=L-\sqrt{L^2-2p_gt}$. For the velocity we find
\begin{equation}
V=\dot\beta={p_g\over\sqrt{L^2-2p_gt}} \ . \label{V=}
\end{equation}
This makes it clear that any genuine
physical determination of the flow (including the flux) requires
boundary conditions at {\em finite} pressures, rather than at $p\to
-\infty$.
\subsection{The class of solutions (27) as a resummation}
Before turning to the discussion of analyticity and conformality
of the solutions, it is important to reflect on the meaning
of these solutions. It is straightforward to verify that if we
start with a Saffman-Taylor solution with width 1/2, that is,
\be
f_0=2t+\ln(e^{\zeta-\hat\zeta}+1) \ , \quad e^{-2\hat \zeta}=1+e^{-2t}
\ee
and perturb it to say $f_0+\epsilon f$, then the equations
of motion are satisfied with errors $O(\epsilon^2)$, for
each
\be
f_n=e^{2nt}(1+e^{-\hat\zeta-\zeta})^n\ , \quad n=1,2,\dots \ , \label{dai1}
\ee
all analytic as Re$\zeta\to \infty$. That is, the basis for perturbations
to $f_0$ grow exponentially in time. We can construct from these,
``center manifold" perturbations that grow slower than exponential
in time, by forming the infinite sum
\begin{eqnarray}
f_a&=&-\sum_1^\infty (-a)^nf_n=\ln(1+ae^{2t}(1+e^{-\hat\zeta-\zeta}))
\nonumber \\&=&\ln(1+ae^{2t})+\ln(1+e^{\zeta_s(t)-\hat\zeta})\nonumber\\
\zeta_s&=&-\hat\zeta-\ln(1+e^{-2t}/a)\ . \label{dai2}
\end{eqnarray}
(For $a=1$, $\zeta_s=\hat\zeta$). Thus, class (\ref{exact})
viewed as $f_0$ plus fluctuations, is simply an infinite resummation
of the unstable manifold spanned by (\ref{dai1}). But this makes
it clear that generic functions are to obtained, reciprocally, by
{\em infinite} summations over class (\ref{exact}), they behaving
very differently from a finite sum. This leaves as open the issue
if conclusions based on class (\ref{exact}) have any unrestricted
applicability. This question is beyond the scope of this paper,
and will be addressed in a future paper.
\section{Analyticity and conformality as a function of time}
\subsection{Solutions of the
equations of motion for the class $g$ and asymptotic stagnation
points}
Considering the solution (\ref{exact}) we
compute
\begin{eqnarray}
f_t &=&\dot\beta-\sum_{k=1}^N{\alpha_k\dot
\zeta_k\over
e^{\zeta-\zeta_k}-1}\label{ft}\\f'&=&1+\sum_{k=1}^N{\alpha_k \over
e^{\zeta-\zeta_k}-1}\label{f'}
\end{eqnarray}
It is useful to consider these equations in various limits. Consider
first $\zeta\to \pm \infty$
on the real axis:
\begin{eqnarray}
\zeta\to \infty:\quad
f_t&=&\dot\beta\quad f'=1 \ , \label{inft}\\\zeta\to -\infty:\quad
f_t&=&\dot\beta+\sum_k\alpha_k\dot \zeta_k\\f'&=&1-\sum_k\alpha_k \
. \label{f'as}
\end{eqnarray}
Substituting in the equation of motion (\ref{lgesym}) we find
\begin{equation}
(2-\sum_k\alpha_k)\beta+\sum_k\alpha_k\zeta_k=2(t-t_0)
\ , \label{sumrule}
\end{equation}
with $t_0$ a constant of integration. Next consider the asymptotic behavior
$\zeta\sim\zeta_k$:
\begin{equation}
\zeta\sim \zeta_k:\quad f_t\sim -{\alpha_k\dot \zeta_k\over
e^{\zeta-\zeta_k}-1}\ , \quad f'\sim
{\alpha_k \over e^{\zeta-\zeta_k}-1}
\end{equation}
Substituting in the equation of motion (\ref{lgesym}), after
dividing by $f'(\zeta)f'(-\zeta)$, we find
\begin{equation}
-\dot \zeta_k f'(-\zeta_k,t)+f_t(-\zeta_k,t)=0 \ .
\end{equation}
(Had we taken the
$\alpha_k$ time dependent the first sub-dominant asymptotic term would
have revealed that indeed the $\alpha_k$ are constant.) Recognizing
that this equation reads $df(-\zeta_k,t)/dt=0$, we introduce the points
$z_k$ in physical space,
\begin{equation}
\bar z_k\equiv f(-\zeta_k,t) \
, \quad \dot z_k=0 \ . \label{stag}
\end{equation}
(Notice that by reflection symmetry $z_k$, as the $\zeta_k$, come
in complex conjugate pairs.) With Re$\zeta_k<0$ we see that $-\zeta_k$
is in the interior of $\Omega$, and therefore $z_k$ is within the
moving fluid. Thus $z_k$
are asymptotic {\em stagnation} points of the flow as the singularities
approach the interface. This result was observed for the first time in
\cite{94DMW}. More correctly, the fluid stagnates $\alpha_k\ln 2$ upstream
from $z_k$ as Re$\zk\to 0$, at $\zeta=i$Im$\bar\zeta_k$.
Finally we rewrite Eq.(\ref{stag}) in the form
\begin{equation}
\bar z_k=\beta(t)-\zeta_k+\sum_l\alpha_l\ln(1-e^{\zeta_l+\zeta_k}) \
.\label{zk}
\end{equation}
\subsection{Necessary Conditions for Asymptotic Analyticity}
The discussion of analyticity is largely
independent of the explicit form of solutions (\ref{exact}). Infinite
channel asymptotics implies that
\begin{equation}
f=\beta+\zeta+F(\zeta,t) \ , \label{genform}
\end{equation}
with the first two terms describing a uniformly translating fluid,
and the last term  its
decoration, (exponentially) vanishing at $+\infty$, so that
\begin{equation}
F,~ F',~F_t\to 0 ~{\rm as~ Re}\zeta\to +\infty \ . \label{FF'Ft}
\end{equation}
Much of the discussion relies just on
this. Moreover, to exhibit the moving singularities, we write
\begin{equation}
F(\zeta,t)=\sum F_k(\zeta-\zeta_k(t))\ ,~~
F_t=-\sum \dot\zeta_k F'_k(\zeta-\zeta_k(t)) \ ,
\end{equation}
with each $F_k$ obeying (\ref{FF'Ft}), and by the analyticity of $f'$
at$-\infty$,
\begin{equation}
F'_k(-\infty) \equiv -\alpha_k \
. \label{defalph}
\end{equation}
Granted this level of detail of $f$'s
form, we have by the equations of motion(\ref{lgesym}), with $f\sim
\beta+\zeta$ as Re$\zeta\to +\infty$,
\begin{equation}
f_t(-\infty)+\dot\beta f'(-\infty)=2 \
,\end{equation}or\begin{equation}(1+f'(-\infty))\dot\beta
+F_t(-\infty)=2 \
,\end{equation}
or
\begin{equation}
(1+f'(-\infty))\dot\beta +\sum \alpha_k\dot \zeta_k=2 \
,\end{equation}
or
\begin{equation}
(1+f'(-\infty))\beta +\sum \alpha_k
\zeta_k=2(t-t_0) \ . \label{t-t0}
\end{equation}
Let us write
\begin{equation}
1+f'(-\infty) = 2-\sum \alpha_k\equiv 2\lambda \
. \label{deflam}
\end{equation}
Should all work well, with no impediment
against all the $\zeta_k$ approaching the interface Re$\zeta=0$ as
$t\to \infty$, then (\ref{t-t0}) implies
\begin{equation}
\beta \sim {t\over \lambda} \ . \label{btlam}
\end{equation}
It is now straightforward to show that the physical case
requires
\begin{equation}
0<\lambda<1 \ , \label{lamcond}
\end{equation}
and hence by (\ref{deflam})
\begin{equation}
|f'(-\infty)|<1 \ . \label{f-inf}
\end{equation}
To see this, let us determine
$\zeta(\zeta_0,t)$ for Re$\zeta\to +\infty$. The differential equation
for $\zeta$, Eq.(\ref{eqzeta}) reads here
\begin{equation}1\sim \zeta_t+\dot \beta \ .
\end{equation}
Integrating
\begin{equation}
\zeta\sim t-\beta+\zeta_0\sim (1-{1\over \lambda})t+\zeta_0 \
. \label{zet}
\end{equation}
A given, far downstream, $\zeta_0$ lies on
a line of constant pressure, imaged by $f$ at $t=0$ into a curve in
$z$-space of constant pressure. Should $\zeta(\zeta_0,t)$ at later times
be {\em further} downstream, then by (\ref{genform}) and (\ref{FF'Ft})
this same fluid particle lies on successively {\em flatter} pressure
curves, so that a flattening profile propagates upstream towards the
interface. Of course, precisely the opposite must occur, and so by
(\ref{zet})
\begin{equation}
1-{1\over \lambda}<0 \ , \quad {\rm or}~0<\lambda<1 \ .
\end{equation}
With $\zeta$ of (\ref{zet}), we now know the trajectories of far
downstream particles:
\begin{equation}
f(\zeta,t)\sim \beta+\zeta\sim t+\zeta_0 \ ,
\end{equation}
just reflecting uniform unit flux of the
fluid. On the other hand, with the interface at Re$\zeta=0$ and
Re$\zeta_k\to 0^-$, $F$ is bounded outside sufficiently small disks
about the $\zeta_k$'s, and so $F(is,t)$ is bounded outside sufficiently
small intervals of $s$. Observing Eq.(\ref{ft}) and realizing that for
long times $\dot\zeta_k\to 0$, (and see a formal proof in the next
section), we reach the conclusion that
\begin{equation}
f(is)\sim\beta\sim {t\over \lambda}
\end{equation}
and so the interface is moving
at velocity $1/\lambda>1$. Conservation of flux then imposes that the
net width of the moving interface is just $\lambda$ times the
full channel width, which to be physical (not to fail $h$'s
conformality) must lie between 0 and 1. It is thus clear that of all
possible $f$'s, only those obeying (\ref{f-inf}) meet physical boundary
conditions. We next show that for just these $f$'s analyticity is never
lost in finite time.
\subsection{Absence of Violations of Analyticity}
We prove now that the solutions do not lose
analyticity so long as (\ref{f-inf}) is obeyed. $f$ can fail to be
analytic in the physical region in two possible ways: either $|\beta|$
diverges, leaving $f$ defined nowhere, or the moving singularities cross
Re$\zeta=0$ into the physical regime. By presumption, $\beta$ is finite
and Re$\zeta_k<0$ at $t=0$, so that until such a disease occurs,
Re$\zeta_k$ are bounded from above. We shall now demonstrate $f$'s
analyticity for all future times by exhausting the possibilities. First,
consider that some of the $\zeta_k$'s tend to $-\infty$ at some finite
or infinite time. Call the set of all such indices $k$ $\tilde S$, and
its complement, $S$, then the set of singularities remaining bounded. We
show $\tilde S$ is empty without proviso. According to the equations of
motion for the singularities, for $\zeta_k\to-\infty$,
by (\ref{genform}) and (\ref{FF'Ft})
\begin{equation}
\bar z_k=f(-\zeta_k)\sim \beta -\zeta_k\ , \Rightarrow~\beta\to
-\infty
\end{equation}
and produces a disease. But then
\begin{equation}
\sum_{\tilde S} \alpha_k\zeta_k\sim \beta
\sum_{\tilde S} \alpha_k+bdd
\end{equation}
and by (\ref{t-t0}) and (\ref{deflam})
\begin{equation}
2t\sim \beta (2-\sum_k \alpha_k)+\beta
\sum_{\tilde S} \alpha_k +bdd=\beta(2-\sum_S\alpha_k)+bdd \
. \label{0}
\end{equation}
Should $2-\sum_S \alpha=0$ this can
(potentially) occur at finite time, else as $t\to \infty$,
$2-\sum_S\alpha <0$ ($\beta\to -\infty$). That is, we
require
\begin{equation}\sum_S \alpha \ge 2 \
. \label{1}
\end{equation}
Now, consider $\zeta_k$ with any $k\in S$. From (\ref{stag}) and
(\ref{genform})
\begin{equation}\bar z_k =\beta -\zeta_k+\sum_\ell
F_\ell(-\zeta_k-\zeta_\ell)\sim\beta+\sum_{\ell\in S}
F_\ell(-\zeta_k-\zeta_\ell) \ ,
\end{equation}
or, with $\beta\to -\infty$
\begin{equation}
\sum_{\ell\in S}
F_\ell(-\zeta_k-\zeta_\ell)\to +\infty, ~{\rm for~each}~k\in S\
. \label{2}
\end{equation}
Eq.(\ref{2}) can be satisfied only if
$\zeta_k+\zeta_\ell\to 0,\pm 2\pi i$ so that $F_\ell$ becomes singular,
i.e. when {\em all} the Re$\zeta_k\to 0^-$, $k\in
S$. Define
\begin{equation}
S_k\equiv \{\ell|\bar \zeta_k+\zeta_\ell \to
0,\pm 2\pi i\} \ . \label{3}
\end{equation}
Clearly, $k\in S_k$, so that no $S_k$ is empty, and the $S_k$'s are a
partition of $S$. Then, (\ref{2}) is
\begin{equation}
\sum_{\ell\in S_{\bar k}}
F_\ell(-\zeta_k-\zeta_\ell)\to +\infty \
. \label{2'}
\end{equation}
Here we use a detail of $F_\ell$ of
(\ref{exact}): the logarithm tends to $-\infty$, and so
(\ref{2'}) requires
\begin{equation}
{\rm Re} \sum_{\ell\in S_{\bar k}}
\alpha_\ell < 0 \ , \label{4}
\end{equation}
where $\zeta_{\bar k}\equiv
\bar \zeta_k$. Summing (\ref{4}) over distinct $S_{\bar k}$, we then
have
\begin{equation}
\sum_S \alpha={\rm Re}\sum_S \alpha<0 \ , \label{5}
\end{equation}
but this contradicts (\ref{1}). Finally should
$S$ be empty, (all $\zeta_k\to -\infty$), (\ref{0}) is $t\sim\beta$,
impossible since $t>0$. We thus conclude that {\em no} $\zeta_k$ can go
to $-\infty$ for any $t>0$. We are now left with the circumstance that
all $\zeta_k$ are bounded. Should $\beta$ be finite, then it is
impossible for any $\zeta_k$ to cross Re$\zeta=0$. This is contingent
upon $F_k$ imaging an arbitrarily small disc about $\zeta_k$ with
arbitrarily large modulus, such as is the case for $F_k$ of
(\ref{exact}) with Re$\alpha_k\ne 0$. In this circumstance, and with
$\beta$ finite, $|f(\zeta_k)|\to
\infty$, $|f(-\bar\zeta_k)|=|z_k|=$finite, and so $\zeta_k$ cannot
approach $-\bar\zeta_k$, i.e. Re$\zeta_k$ cannot approach 0. Accordingly
the discussion has contracted to Re$\zeta_k\to 0$, $|\beta|\to\infty$,
for which (\ref{0}) reads
\begin{equation}
2t\sim \beta(2-\sum \alpha_k)=\beta(1+f'(-\infty)) \ . \label{6}
\end{equation}
Should $f'(-\infty)=-1$, then $\beta$ can diverge at finite $t$, and generally
a finite-time loss of analyticity can occur. Otherwise we have just
$t\to +\infty$ , and so,
\begin{equation}
f'(-\infty)<-1 \ , \quad \beta \to -\infty \ , \label{7}
\end{equation}
or
\begin{equation}
f'(-\infty)>-1 \ , \quad \beta\to +\infty \ . \label{7'}
\end{equation}
(Since $|\beta|\to
\infty$, {\em every} $\zeta_k$ crosses Re$\zeta=0$ ``simultaneously"
at $t=+\infty$.) Should case (\ref{7}) hold the argument leading to
(\ref{5}) is appropriate, save that $S$ includes {\em all} the
$\zeta_k$'s, and so $\sum \alpha_k<0$, i.e., $f'(-\infty)>1$, a
contradiction to (\ref{7}). Thus, only case (\ref{7'}) remains, in
which case, the same argument leading to (\ref{5}) is now
$\sum\alpha_k>0$, i.e. $f'(-\infty)<1$, which with (\ref{7'}) is
$|f'(-\infty)|<1$. To summarize what we have now demonstrated,
\vskip 0.5 cm {\em Observation 1}: Unless $f'(-\infty)=-1$, then $f$ remains
analytic throughout the physical region for all positive time. $t$ can
only continue to $+\infty$ if $|f'(-\infty)|<1$.
\vskip 0.5 cm
The curious part of this result is that if $|f'(-\infty)|>1$, then
$f$ remains analytic, but the $\zeta_k$ never get to cross
Re$\zeta$=0. This happens because then $f$ must lose {\em conformality}
because $f_t$ has diverged (i.e. $\dot \beta$ and $\dot\zeta_k$'s become
infinite) although $f$ is finite and analytic at such an
instant. Should $f_t$ diverge, then the equations of motion imply that
$f'=0$ at some point on the interface, i.e. that $v\to \infty$ at some
point of the physical fluid. What our observation says is there must surely
be finite-time singularities should $|f'(-\infty)|>1$. These
singularities represent precisely an impending violation of {\em
boundary} conditions. To see this, recall that our wall
boundary conditions are $x_s(p,s)=0$ on $s=0,\pi$. By reflection
symmetry, $f$ has been analytically continued to $-\pi<s<\pi$, rather
than just the physical $0<s<\pi$. However such an $f$ does
not necessarily map $s>0$ to {\em  just} $y>0$. It will do so provided
that $f$ is conformal. Should $f$ not be a contraction at $-\infty$, it
is inevitable that, as the singularities approach the interface, $f$
will begin to map across half channels, and so, such an $f$ fails
physical boundary conditions, and so is not an admissible solution. (To
see this, by Eq.(\ref{uw}), $g'/g=f'/u$, or $g\sim u^{f'(-\infty)}$ as
$u\to 0$- i.e. everywhere within the radius of that $a_k(t)$ with
smallest modulus. That is,with arg$u=\phi$, arg$w\sim
f'(-\infty)\phi$, and exceeds half channel width unless
$|f'(-\infty)|<1$.) Consequently our observation reads that all
admissible solutions under physical channel boundary conditions are
free from finite time failures of analyticity, so that the conformal
machinery we are employing is perfectly applicable. The difficulty is,
while $|f'(-\infty)|>1$ always leads to finite time failures in $f$'s
conformality, the same disease can arise for admissible solutions as
well. This difficulty is profound, and we will expand upon it in the
next section.
\section{The asymptotics and the emergence of One Finger}
\begin{figure}
\hskip 0.1cm
\epsfxsize=8.0truecm
\epsfbox{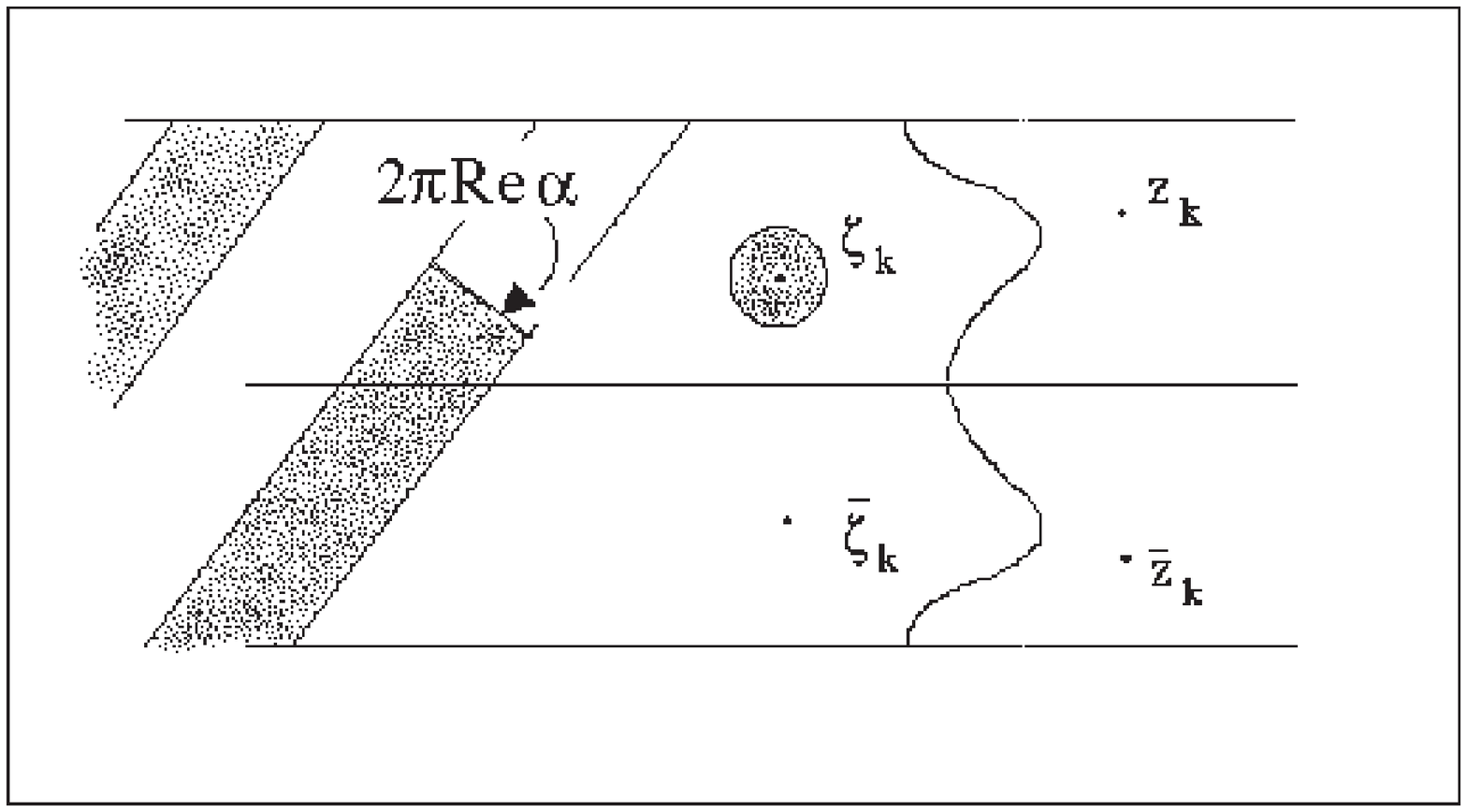}
\vskip 0.2cm
\caption{The mapping of a neighborhood of one of the singularities}
\label{tilt}
\end{figure}

We are ready now to establish a significant
result: For $t\to \infty$ the physical channel supports one
finger. In the doubled channel, that is considered in all the literature,
this of course means two fingers with a
stagnation point on the symmetry line. In particular, one finger
in this geometry is a physically incorrect result. Mathematically this is
equivalent to the following
\vskip 0.5 cm
{\em Observation 2}: For solutions that do not lose conformality
for $t\to\infty$, and for which all $\alpha_k$ are generic with
Im$\alpha_k\ne 0$,
all $\zeta_k\to 0,\pm i\pi$.
\vskip 0.5 cm
{\em Demonstration}: Consider one of the terms in the sum (\ref{exact}), say
$\alpha_k\ln\left(1-e^{\zeta_k-\zeta}\right)$. Consider a circle of
radius $|\epsilon|$ around the singularity, $\zeta=\zeta_k+\epsilon$,
where $\epsilon\equiv |\epsilon|e^{i\phi}$, $|\epsilon|$ is small, and
$-\pi<\phi<\pi$. We investigate the image of this small circle under
$f$. For small $|\epsilon|$
\begin{equation}
f\sim \beta(t)+\alpha_k(\ln|\epsilon|+i\phi)~ {\rm mod}\pm i\pi \
.\label{image}
\end{equation}
For a given $|\epsilon|$ this image is a line of length $2\pi|\alpha_k|$
perpendicular to the direction of $\alpha_k$ (see Fig. 3). Consider next
the punctured
disk around the singularity with radius $|\epsilon|$. This disk
is imaged onto a series of slanted strips ordered along the direction
of $\alpha_k$ with the above width as shown in Fig.\ref{tilt}.
\narrowtext
As time progresses and $\beta$ increases, this series moves to the right,
entering the physical domain. (Remember, this is a neighborhood of a
singularity, not a singularity!). Eventually, when $|$2Re$\zeta_k|<|\epsilon|$
the image of a very small $|\epsilon|$-neighborhood of $\zeta_k$ must include
the stagnation point $z_k$, the fixed image of $-\bar\zeta_k$, and must for
all future
time always include it. But as $t$ increases, the fixed width strip containing
$z_k$ is moving arbitrarily far to the right. This would seem impossible. The
only way to resolve this conundrum is that
\begin{equation}
\zeta_k(t) \to 0~{\rm or}~\pm i\pi\quad {\rm
as}~t\to\infty \ . \label{zap}
\end{equation}
When this happens $e^{\zeta_k}\to e^{\bar\zeta_k}$ and $\bar \zeta_k$ (mod
$2\pi i$
if necessary) is also in the $|\epsilon|$-neighborhood, and its term in
(\ref{exact})
must be also included in (\ref{image}). But
\begin{equation}
\alpha\ln(1-e^{\zeta_k-\zeta})+\bar\alpha
\ln(1-e^{\bar\zeta_k-\zeta})\to
2{\rm Re}\alpha\ln(1-e^{\zeta_k-\zeta}) \
. \label{noslant}
\end{equation}
and so the strips begin to rotate to the horizontal, and sufficiently so
that $z_k$ always remains within its strip. (\ref{zap}) is the only way to
have
Re$\zeta_k\to 0$.

It should be stressed that even though an asymptotic finger solution
is emerging, its width is in no way selected. Moreover, over the
duration of an actual experiment, not {\em all} the singularities
need be yet in asymptotic proximity to Re$\zeta=0$. That is, a subset of
$\zeta_k$'s of the same order of magnitude will carry out migration
toward 0 or $i\pi$, while other sets with extremely large -Re$\zeta_k'$
are still very far from asymptotic behavior.

To see this, consider as usual $f$ of (27) built from those $\zk$
that are sufficiently close to one another, and consider an
exponentially small perturbation, built from $\zk'$ with
Re$\zk'\le -x\ll$ Re$\zk$, $x\gg 1$:
\be
f=\beta+\zeta+\sum \alpha_k\ln (1-e^{\zk-\zeta})+
\sum \alpha_k'\ln (1-e^{\zk'-\zeta}) \ . \label{sof1'}
\ee
Clearly, near the interface the last term is $O(e^{-x})$.  Writing the
equations for $\zk'$,
\be
\bar z_k'=\beta-\zeta_k'+O(e^{-x})\ , \quad
\rightarrow {\rm Re}\bar z_k'\ge\beta+x \ . \label{sof2}
\ee
Writing the $\beta$ equation
\begin{eqnarray}
(1-\sum\alpha_k)\beta&+&\beta+\sum \alpha_k\zk
\nonumber\\&+&(\sum\alpha_k'
\zk'-\beta\sum\alpha_k')=2t-k\ . \label{sof3}
\end{eqnarray}
Multiplying each equation of (\ref{sof2}) by the $\alpha_k'$ and
summing, we see, with $k=\sum\alpha_k'\bar z_k'$,
\be
(1-\sum\alpha_k)\beta+\beta+\sum \alpha_k\zeta_k=2t+O(e^{-x}) \ , \label{sof4}
\ee
while the equations for each $\zk$ are
\be
\bar z_k=\beta-\zk+\sum\alpha_\ell\ln(1-e^{\zk+\zeta_\ell})+O(e^{-x})
\ . \label{sof5}
\ee
We see then, by (\ref{sof4}) and (\ref{sof5}) that the near
set of $\zk$'s, for $x$ large enough, behaves exactly as a full system,
and forms a finger of width $\lambda=1-\sum\alpha_k/2$, and $\beta=t/\lambda$.
But then, by (\ref{sof2}), for $t\le \lambda x/2$, Re$\zk'<-x/2$, and the
$\zk'$
still play an exponentially small role up to very long times during which
a $\lambda$-finger is propagating. Ultimately for $t\sim \lambda x$, the
$\zk'$ become asymptotic, and the finger metamorphoses into a
$\lambda'=1-(\sum\alpha_k+\sum\alpha_k')/2$. So, from a physical
viewpoint, with $\sum \alpha_k+\sum\alpha_k'=1$, the finger will become
$\lambda=1/2$ long after the experiment is over. But $f'(-\infty)=
1-\sum\alpha-\sum\alpha'=0$ no matter how large $x$ may be. We see from
this that $f'(-\infty)=0$ is insufficient to determine $\lambda=1/2$.
The width is determined just by the $\zk'$s near enough to Re$\zeta=0$. Indeed,
in (\ref{sof1'}) the $\sum\alpha_k'$ can be chosen arbitrarily
with no consequence during the experiment. Note that (\ref{sof1'})
is the transparent implementation of the manipulations of \cite{98Min}. It
is clear that \cite{98Min} then has no significance for selection of a 1/2
finger. (By choosing the $\sum\alpha'$ arbirtrarily, the argument of
\cite{98Min} would then show the ``selection" of any $\lambda$ whatever.)
These comments pose a limitation on Observation 2 as well: The
larger the number of $\zeta_k$'s we choose, the more semi-asymptotic
regimes the initial conditions can be chosen to determine, so that in the limit
as this number diverges, no final asymptotics need exist.

\section{Violations of conformality: finite time singularities}
The discussion that follows critically assumes that $f$ is in the
manifold (\ref{exact}).
Should the velocity of a fluid point diverge, $h'\to \infty$ and
correspondingly $f'\to 0$, and so the map $f$, while remaining
analytic, has locally lost conformality. As we observed at the end
of Sect. 3.c, should $|f'(-\infty)|>1$, for infinite channel flow,
then there {\em must} occur a violation of conformality. However,
this circumstance is trivial, in that we realized the failure occurred on the
walls of the channel, and hence is not a solution under our boundary
conditions. Regrettably, this is far from the only way in which such
violations occur. However, it is not too hard to determine when $f'(\zeta)=0$
for some $\zeta$ within the physical fluid. Since for infinite channel
flow $f'(\infty)=1$, and so long as $f'(\zeta)\ne 0$ for Re$\zeta\ge 0$,
then $|f'(\zeta)|>0$ throughout the physical region. But then $|f'|$,
since $f'$ is analytic, has its minimum value over a region on the boundary of
that region, which now evidently means the interface at Re$\zeta$=0. It
now follows that if a failure of conformality occurs, it must first
appear on the interface, and so the condition for a finite-time
singularity is
\begin{equation}
f'(is)=0
\end{equation}
for some $0<s<\pi$ (but not a boundary violation at 0 or $\pi$).

To understand what happens, consider the behavior of the Saffman-
Taylor solutions, that is $N=1$ in (\ref{exact}). By (\ref{sumrule})
\begin{equation}
\beta={2t\over 2-\alpha}-{\alpha \rho_0\over 2-\alpha} \ , \label{c.1}
\end{equation}
where the one singularity, $\rho_0$, is real as is its corresponding
$\alpha$. Then,
by (\ref{zk}), with $z_0=0$,
\begin{equation}0={2(t-\rho_0)\over 2-\alpha}+\alpha\ln
(1-e^{2\rho_0})
\end{equation}
or
\begin{equation}
t=\rho_0-{1\over 2}\alpha(2-\alpha)\ln(1-e^{2\rho_0}) \
. \label{c.2}
\end{equation}
(\ref{c.2}) is soluble for $\rho_0$ as long
as $t'(\rho_0)\equiv dt/d\rho_0\ne 0$. For $\rho_0\to
-\infty$, evidently $t\sim \rho_0$ and $t$ can increase from the far
past. For $t'=0$, generally, $t$ will have then a maximum, and so $\dot
\rho_0\to \infty$, inducing a finite
time singularity. But
\begin{equation}t'=0=1+{\alpha(2-\alpha)\over e^{-2\rho_0}-1} \ ,
\end{equation}
or
\begin{equation}
e^{-2\rho_0}=(1-\alpha)^2
=[f'(-\infty)]^2\ , \quad e^{-\rho_0}=|f'(-\infty)| \ . \label{c.3}
\end{equation}
We now see that with $\rho_0<0$, $t'=0$ is
impossible for $|f'(-\infty)|<1$, but certain otherwise. That is, time
``locks" for the inadmissible cases, and only these cases. With $\dot
\rho \to \infty$ (but f finite and analytic), $f_t\to \infty$, and
we seek a $\zeta$ on the interface where $f'(\zeta)=0$. But
\begin{equation}f'=0=1+{\alpha \over e^{\zeta-\rho_0}-1} \ ,
\end{equation}
or
\begin{equation}
e^\zeta=(1-\alpha)e^{\rho_0}={f'(-\infty)\over
|f'(-\infty)|}=\pm 1 \ .
\end{equation}
This is the general nature of
the failure of conformality. At some time, $t_0$, time locks and various
of $\dot \beta$ and $\dot\zeta_k$ diverge, and hence $f_t$
diverges, although $f$ can be perfectly finite. Simply, the system
(\ref{sumrule}),(\ref{zk}) becomes locally non-invertible for the
$\zeta_k$'s.

We noticed in Sect.2.F that Re$f'(\zeta)=\lambda$ for our
translating solutions so that $|f'(\zeta)|=\lambda>0$, and the interface
is conformal, and any violation of conformality is the interior
representation of sinks or sources. But Re$f'(is)=y'(s)$, or
$y=\lambda s$, and so $x(s)=X(y)$, and the interface is a graph
of $x$ on $y$. Generally, a graph with finite $X'(y)$ (i.e. a differentiable
graph) won't fail conformality.

For the class of solutions in Sect 2.E
 with all $\alpha_k>0$, it is easy to see that
$y'(s)>\lambda$, and so $y$ is monotone in $s$, and hence the solution
is a graph with $|f'(\zeta)|>\lambda$ and so always conformal for
$0<\lambda<1$. The only possibility for a failure of conformality
is with complex $\alpha$'s and the interface, a graph in the
far past, about to become not a graph. (Indeed, the generic rotation
mechanism for the emergence of one asymptotic finger is
Im$\alpha_k \ne 0$ for all $k$.) Real time singularities thus can arise
when a ``balloon" (not a graph) is about to form.

With $\alpha_k=a_k+ib_k$, Re$f'(is)$ is modified with $b_k\ne 0$ by terms
exponentially small when the singularities are far from Re$\zeta$=0, that is
$\sim e^{{\rm Re}\zeta_k}$. Thus conformality can fail only when
singularities enter the rotation mechanism of Sect.4, which turns
fingers into balloons. In particular, this is definitely beyond
the perturbative regime, and when the nonlinearities have become very strong.
The usual linear stability analysis, with temporal exponents proportional
to wavelength, simply means that fluctuations very rapidly bring the
solution into the strongly nonlinear regime. Indeed, rather than
infinitely wrinkled, distorted interfaces, if the rotation mechanism can work,
a smooth single balloon is the consequence of the nonlinearities, {\em
provided}
class (\ref{exact}) obtains. We will have more to say about this in
\cite{99Fei},
as it transpires that the unstable bhavior of the infinite channel is
physically
significantly wrong. Any
inspection of early interface structure - say in Saffman-Taylor's
original paper - reveals that it is balloons that are pervasive, and not
graphs or fingers. Let us consider the simplest balloon.

With the channel-doubling conjugacy of Sect. 2.F implicit,
consider the solutions with one $\zeta_k$:
\begin{eqnarray}
f&=&\beta+\zeta+\alpha\ln(1-e^{\zeta_0-\zeta})+\bar\alpha
\ln(1-e^{\bar\zeta_0-\zeta})\nonumber\\
\alpha&=&a+ib;~~a,b>0;~~\zeta_0\equiv -\xi+i\eta,\xi>0 \ . \label{n1}
\end{eqnarray}
The equation of motion $z_0=f(-\bar\zeta_0)$ is
\begin{eqnarray}
z_0&=&\beta-\bar\zeta_0+\alpha\ln(1-e^{2{\rm Re}\zeta_0})+\bar\alpha
\ln(1-e^{2\bar\zeta_0})\nonumber\\
z_0&=&\beta+\xi+i\eta+\alpha\ln(1-e^{-2\xi})+\bar\alpha
\ln(1-e^{-2\xi-2i\eta})\nonumber \ .
\end{eqnarray}
Its imaginary part is
\begin{equation}
y_0=\eta+a{\rm Arg}(e^{2\xi}-e^{-2i\eta})-\frac{b}{2}\ln
{|e^{2\xi}-e^{-2i\eta}|^2\over (e^{2\xi}-1)^2} \nonumber \ ,
\end{equation}
or
\begin{equation}
y_0=\eta+a\tan^{-1}{\sin 2\eta\over e^{2\xi}-\cos 2\eta}
-\frac{b}{2}\ln(1+{\sin^2\eta\over \sinh^2\xi}) \ . \label{n2}
\end{equation}
with the usual principal value of $\tan^{-1}$ correct. $y_0$ is here
half the distance between the two ``stagnation" points $z_0$ and
$\bar z_0$. The level curves of the RHS of (\ref{n2}) with $0\le y_0
\le \pi$ are precisely, for each $y_0$, the trajectory curve of a
$\zeta_0$, however it be parametrized by $t$. There are three types
of trajectories that connect to $\xi\to \infty$ (with $y\to y_0$ as
$\xi\to \infty$):\\
(i) $\tan^{-1} \frac{a}{b}<y_0\le \pi$: the trajectory monotonically
(in -$\xi$) increases from $y_0$ to $\pi$ as $\xi\to 0$.\\
(ii) $a\tan^{-1} \frac{a}{b} -\frac{b}{2}\ln(1+\frac{a^2}{b^2})<y_0
<\tan^{-1}\frac{a}{b}$:
the trajectory moves from $y_0$ to $\pi$ as $\xi\to 0$, initially to lower
$\eta$ values, and with a unique minimum.\\
(iii) $0\le y_0<a \tan^{-1} \frac{a}{b}-b\ln(1+\frac{a^2}{b^2})$:
the trajectory monotonically flows from $y_0$ to $0$ as $\xi\to \infty$.

Trajectories of types (i) and (ii) rotate to $\pi$, and the ``walls"
at $\pm \pi$ are closed to flow, with a balloon symmetric about
$y=0$ moving down the channel. Type (iii) has $\zeta_0$ and $\bar \zeta_0$
both rotate to $\eta=0$, blocking flow along $y=0$, with fluid advancing
along the $\pm \pi$ walls. By Sect.2.F, (i) and (ii) have blocked flow at
$y=0,\pi$ with the balloon symmetric about $y=\pi/2$, while (iii)
has flow blocked along $\pi/2$, since the upper half poles are both
rotating together to $\pi/2$. This is unphysical and non-generic: the
rotation mechanism of Sect.4 can only lead to this under extra,
nonphysical, symmetry, which of course is exactly what the method
of Sect.2F creates.
With two {\em generic} poles, case (iii) would not have occurred.
Regrettably, this generic version is not analytically tractable.

{\em However}, while it turns out that type (iii) never encounters finite
time singularities, not all
the ``good" types, (i) and (ii) trajectories are free of disease. That is,
there is, for each $a$ and $b$, a minimum gap, $2y_0^{\rm min}$ between the
stagnation points that allows the interface to squeeze down through the gap,
and then re-emerge, blooming out into a balloon. For any
smaller $y_0$, the interface is squeezed into a cusp, unable to pass through
the gap without penalty of a singularity. This attempt to squeeze through
and balloon out is the generic disease that $\sigma=0$ theory is plagued
by: there are
a fraction of initial conditions that fail. In fact, this is precisely
where surface tension needs to be enlisted. With surface tension
the stagnation points are no longer constants of the motion, and
indeed will move apart just enough to allow the incipient balloon to pass
through the gap. It is noteworthy that in experimental studies
such a phenomenon always appears at the initiation of flow (cf.
\cite{58ST,87TZL}).
\section{Finiteness}
So far we have considered a channel filled with fluid infinitely far
downstream.
This is of course unphysical. Any experimental apparatus introduces by
necessity
some additional boundary condition on the physical
fluid far downstream, requiring mathematical
boundary conditions to model this termination.

We recall that the possibility of adding a sink located at some finite
position was discussed in Sect. 2F. We can think of other ways
to have the fluid itself finite.  First, consider the idealized Hele-Shaw cell.
 At
a long
distance downstream we erect
a baffle cross-wise to the channel -- say at $x=0$. Behind the baffle we
have a pump
controlled to maintain an exactly constant unit flux of fluid through the
baffle.
With a uniform enough baffle, we have $\B.v(0,y,t)\propto \hat x$ as an
approximate
boundary condition. Thus $v_y=0=-\partial_yp$ on $x=0$ or $p(0,y,t)\equiv
-p_g(t)$,
$p_g$ the positive gauge pressure on the finite fluid from the interface at
$p=0$
to the baffle. That is
\begin{equation}
{\rm Re} f=0~~{\rm at~Re}\zeta=\frac{p_g(t)}{V}\equiv \xi_g(t) \ .\label{vn1}
\end{equation}
But with $f$ reflection symmetric, we have
\begin{equation}
f(\zeta,t)+f(\bar\zeta,t)=0\quad {\rm at}~ \zeta+\bar\zeta=2\xi_g(t) \ .
\label{vn2}
\end{equation}
Assuming the fluid is analytic over any region containing Re$\zeta= \xi_g$
in its interior,
we then have by analytic continuation
\begin{equation}
f(\zeta)+f(2\xi _g-\zeta)\equiv 0 \ , \label{vn3}
\end{equation}
for all $\zeta$ in the region of analyticity. This exposes the real power
of reflection
symmetry: not only is there a relation of the upper physical channel to the
lower unphysical
one, but under finite boundary conditions from very high pressures to very
low ones. In this case there
is no full exponential decoupling of efflux from interface motion. This is
precisely the ``enforced
symmetry" between $\zeta$ and $-\zeta$ mused about in the sink solution of
(\ref{nss}) with
$\alpha=1$ in Sect.2F with its upstream pointing 1/2 Saffman-Taylor finger.
We will explore
this momentarily, after discussing the variant to Hele-Shaw, and a related
other pair
of terminations.

An obvious variant to fixed velocity on the cross-channel line at $x=0$ is
to simply
open (cut off the end of) the channel, so that $p(0,y,t)\equiv
p_a=$const=atmospheric
pressure. We then have
\begin{equation}
{\rm Re} f=0~{\rm at}~{\rm Re}\zeta=-\frac{p_a}{V(t)} =
+\frac{p_g}{V(t)}\equiv \xi_g(t)>0
\ . \label{vn1'}
\end{equation}
Just as before, we now have
\begin{equation}
f(\zeta)+f(2\xi_g-\zeta)=0\ , \label{vn3'}
\end{equation}
so that both variants entail the identical calculations, save for the
driving fluxes:
\begin{equation}
2=f'(\zeta)f_t(-\zeta)+f'(-\zeta)f_t(\zeta) \ , \label{vn4}
\end{equation}
in the Hele-Shaw case, whereas
\begin{equation}
2\to 2V(t)=\frac{2p_g}{\xi_g(t)} \ , \label{vn4'}
\end{equation}
in the constant pressure termination, ultimately determining the non-steady
$V(t)$ in this
case, as we saw in the most elementary versions of (\ref{vn3'}) with
$f=\beta+\zeta$ in
(\ref{bet+zet})-(\ref{V=}) of 2F.

The other pair of variants replace the cross-channel
line at $x=0$ with a small circular aperture of radius $a$ all along which
either
$v_r\equiv -1/2a$ so that $V=1$ by (\ref{intcond}), or again $p=p_a$ and
Re$\zeta=\xi_g(t)$.
These circular aperture problems are mathematically related by
exponentiation to the cross-
channel line versions, and technically much harder to discuss with closed
solutions.
However with $p=p_a$ on the circular aperture, there must be a singularity
in the interior
of the aperture to sink the full-flux that must enter it if we seal off the
channel
arbitrarily far downstream, so that all fluid must efflux through the
aperture, and in
this case, flow stagnates far to the right, and so whatever we do far
enough to the
right will indeed be exponentially suppressed. If the singularity is just a
simple pole,
then it is a sink, generally moving within the interior of the aperture. By
circle symmetry,
the analogue for (\ref{vn1'}) is the fluid gathering into a moving sink to
right of $x=0$, rather
than becoming flat at infinity. For example, fluid with surface tension
after emerging
from the shaping channel would form
a vena contracta, and so, reminiscent of a moving sink. This was the physical
motivation of our consideration of (\ref{nss}) with $\alpha=1$ in Sect.2.F.

Let now attempt to solve for an $f$ obeying (\ref{vn3}) or (\ref{vn3'}).
Setting $\zeta\to
\zeta+\xi_g$, (\ref{vn3'}) is
\begin{equation}
f(\xi_g+\zeta)=-f(\xi_g-\zeta)\ . \label{vn3''}
\end{equation}
It is easy to check by direct substitution that
\begin{equation}
f(\zeta)=A(\zeta-\xi_g)-A(-\zeta+\xi_g) \ , \label{vn5'}
\end{equation}
with $A$ arbitrary. Consider
\begin{equation}
A(\zeta)=\frac{\zeta}{2}+\sum\alpha_k\ln(1-e^{\zeta_k-\zeta}) \ , \label{vn6}
\end{equation}
and so,
\begin{eqnarray}
f(\zeta)=\zeta-\xi_g(t)&+&\sum\alpha_k\ln(1-e^{\zeta_k+\xi_g-\zeta})\nonumber\\&
-&
\sum\alpha_k\ln(1-e^{\zeta_k-\xi_g+\zeta}) \ . \label{vn7}
\end{eqnarray}
Eq.(\ref{vn7}) is the entire class (\ref{exact}) of solutions meeting our
boundary requirements
$(\beta=-\xi_g)$.

As a first example, consider just one $\alpha$ and choose
$\zeta_0=-\xi_g+i\pi$.
(\ref{vn7}) then is
\begin{equation}
f(\zeta)=\zeta-\xi_g+\alpha\ln(1+e^{-\zeta})-\alpha\ln(1+e^{-2\xi_g+\zeta})
\ . \label{vn8}
\end{equation}
For $\xi_g\ll 1$ this solution is a single Saffman-Taylor finger with an
arbitrary
width. We insist however that there be no flux going off to infinity,
in fact no flux for Re$\z>2\xi_g$ for a fully pinched vena contracta.
(This would have been automatic in the case of the circular aperture.)
To sink all flux requires $\alpha=1$ (cf.
the discussion after Eq.(\ref{nss})), and so
\be
f(\z)=\z-\xi_g+\ln(1+e^{-\z})-\ln(1+e^{-2\xi_g+\z})\ . \label{vn8'}
\ee
which for $\xi_g\gg 1$ is precisely a $\lambda=1/2$ Saffman-Taylor finger
(\ref{10}).
This is our first piece of evidence that $\lambda=1/2$ is connected to
finiteness.

But, all is not well. Notice that
\be
f'=\frac{1}{1+e^{-\z}}-\frac{1}{1+e^{2\xi_g-\z}} \rightarrow f'(-\infty)=0,
\ee
and
\be
f'(+\infty)=1-1=0 \ .
\ee
But then, unless $\dot \xi_g$ is always infinite, (\ref{vn4}) and
(\ref{vn4'}) are only
compatible with $V(t)\equiv 0$, and so these solutions are purely static
and not what
we seek.

Consider then more $\alpha_k$. By Eq.(\ref{vn3'}) if $\zk$ is a singularity
of $f$, so
too is $2\xi_g-\zk$. But then, each asymptotic stagnation point condition
(\ref{stag})
becomes two conditions:
\be
f(-\zk)=\bar z_k,\quad {\rm and}~f(\zk-2\xi_g)=\bar z'_k \ . \label{vn9}
\ee
Together with the $\zeta\to \infty$ equation for $\beta$, there are about
twice as many equations as variables
unless $\xi_g=0$, in which case $p_g=0$, and there is no motion.
We have already seen
that one real $\zeta$ has no flux, and it is reasonable clear
that all other cases entailing too many equations are inconsistently
over-determined. By (\ref{vn3''}) $f'(-\infty)=0\Rightarrow f'(\infty)=0$, and
so there can never be flux with $\lambda=1/2$. The above comments of
over-determination hold for all $\lambda$.  That is,
our first four schemes of finite termination allow no motion for $f$'s
of class
(\ref{exact}) with any {\em finite} number of singularities.

On physical grounds, the fluid emerging into atmosphere becomes
3-dimensional, and the
derivation of Darcy's law breaks down. Eq. (\ref{vn1'}) must be too stringent.
Equivalently, it is not feasible to have a baffle with $\B.v\propto \hat x$
all along
its length. To the contrary, we easily imagine fluid racing vastly
faster through some holes in the baffle rather than others; this choice can
readily
vary in time under minor perturbations of the pump action, etc. So,
there are hosts of singularities very close to the line
Re$z=0$, and
(\ref{vn3}) fails for failure of analytic continuation. (This is most
probably an
over-exaggeration: it seems not necessary that Re$z=0$ is truly a natural
boundary.) On reflection, these comments imply that the physical experiments
that have been performed contain {\em dynamically} determined analyticities,
and so are incompletely posed boundary data configurations.

Let us consider a fifth scheme of finiteness, of a totally different
character from the
previous four. Consider an infinitely long channel, only partly filled with
a finite
body of fluid, with $p\equiv 0$ on the left driven face, and $p\equiv
p_1=-p_g$, $p_g>0$ on the
other, right, free interface. The equation of motion for $f$ on the left
face, are as
usual (\ref{lgesym}) with $\dot V$ surely non-zero:
\begin{equation}
2V(t) =f'(\zeta)f_t(-\zeta)+f'(-\zeta)f_t(\zeta) \ . \label{vn10}
\end{equation}
The right interface lies at
\begin{equation}
{\rm Re} \zeta=\frac{p_g}{V(t)}=\xi_g(t) \ , \label{vn11}
\end{equation}
and so, with $\zeta\to \zeta(\zeta_0,t)$, $\zeta_0$ Lagrangian coordinates,
with free
interface transported to itself,
\begin{equation}
{\rm Re} \zeta_t =\dot \xi_g \ , \label{vn11'}
\end{equation}
and so by (\ref{eqzeta})
\begin{equation}
V(t) =|f'|^2\dot \xi_g+{\rm Re}\bar f'f_t\quad {\rm on}~
\zeta+\bar\zeta=2\xi_g(t) \ . \label{vn12}
\end{equation}
By reflection symmetry, we then obtain a second field equation
in consequence of the second free interface:
\begin{eqnarray}
2V(t)=2\dot
\xi_gf'(\zeta)f'(2\xi_g-\zeta)&+&f'(\zeta)f_t(2\xi_g-\zeta)\nonumber
\\&+&f'(2\xi_g-\zeta)f_t(\zeta)
\ . \label{vn10'}
\end{eqnarray}
The fluid must now simultaneously obey both pde's, (\ref{vn10}) and
(\ref{vn10'}).
It is unquestionably true that this system {\em must} have solutions of a
physical
character, as otherwise the entire 2-d theory should have to be discarded:
This fifth version of
finiteness is {\em entirely} well-posed within a conformal 2-d context.
Singularity
structure, of course, is
 more subtle than our considerations so far, but nevertheless if the right
interface needs
to be a natural boundary (i.e. no further analytic continuation
possible), it is surely the case that so too must be the
left, because the physics at both are  identical.

Let us now consider a class (\ref{exact}) solution. (Imagine the
fluid initially in such a state of perfect repose that its $f$ can
be naturally analytically continued to $+\infty$.) In this case we can take the
limit of (\ref{vn10})
and (\ref{vn10'}) as Re$\zeta\to \infty$. We then deduce that
\begin{equation}
\dot\xi_g f'(+\infty)f'(-\infty)=0 \ . \label{vn13}
\end{equation}
But $f'(+\infty)=1$ (class (\ref{exact})) and $\dot V\ne 0\rightarrow
\dot\xi_g\ne 0$,
and so we have class (\ref{exact}) with
\begin{equation}
f'(+\infty)=1, \quad f'(-\infty)=0 \ , \label{vn14}
\end{equation}
and so {\em only} $\lambda=1/2$ solutions.

Suffice it to say that the equations of motion do possess a well-behaved
solution and
we may conclude that in the only well-posed 2-D finite system we can
construct, finiteness alone determines pattern selection.
\section{Discussion}

We have returned to the Saffman-Taylor problem with the viewpoint
of it as a dynamical system in order to better understand the evolution
of its solutions. In doing so we have carefully re-thought the relevant
boundary
geometry and conditions and realized that reflection symmetry rather than
periodicity is to be imposed. This led to two significant consequences.

First, reflection symmetry and analytic continuation naturally promoted the
equations of motion from a relation pertaining purely to the interface, to
one of a field character throughout the fluid. In consequence, we need
never consider the usual Hilbert transform boundary methods, instead
directly, and largely
algebraically, obtaining solutions and their dynamics.

Secondly, the fluid equations naturally link $f(\zeta)$ and $f(-\zeta)$, so
that
very far downstream details of termination potentially couple to the very far
upstream (above physical fluid) details, such as the singularities
determining the
flow. This counter-intuitive failure of termination details to exponenitally
decouple from the behavior of the interface propelled us to contemplate that
finiteness in this problem is apt to be a deeply significant ``singular"
perturbation
upon the ``physics" of the infinite channel problem. In consequence, we
formulated
the theory from the beginning to include the possibility of variable flux,
a necessity of finite configurations.

Employing our reflection-symmetric field equations, we readily produced a
variety
of elementary solutions and then the pole-dynamics family (\ref{exact}). In
particular we
determined the general form of all translation-invariant solutions, and the
simplest
pole-type solutions more complicated than the original Saffman-Taylor class.
These are characterized by a downstream sink, which when fully sinking all
flux,
produces an upstream pointing 1/2 finger surrounding the zone of efflux.
Considering how the
nature of this finger is contingent upon final termination (a {\em full}
sink at finite
distance), and considering the $\pm \zeta$ symmetry of the equations of
motion, it is
impossible not to wonder that we might be touching upon the origin of
pattern selection.
In this context the reader should not be troubled by the circumstance that
the sink
is now within the body of physical fluid. He should not (or should) because
this is
identical to the situation in the usual infinite channel  flow, when the
full sink
instead of appearing at finite Re$\zeta_s$ is at Re$\zeta_s=+\infty$, still
fully within
physical fluid. (One might think of a M\"obius transformation rotating the
point at infinity
to a proximate point.) This, in fact, should trouble the reader, because it
means that efflux
in the usual case (Re$\zeta_s=+\infty$) has not been physically treated:
The volume of fluid
is conserved only because $\infty-1=\infty$. As we shall see in
\cite{99Fei}, a full
treatment of all real fluid has significant physical consequences. In
particular, it
transpires that each unstable mode $f_n$ of (\ref{dai1}) requires power
from the energetic
sources driving the flow, so that under pump control, the exponentially
growing modes are
sharply supressed, leaving behind, at best, resummations such as class
(\ref{exact}).

We proceeded to analyze the evolution of an arbitray flow, although largely
within the context
of class (\ref{exact}), to better understand how well-formualted the theory
is, and some
general boundary violating circumstances of finite-time singularities -
namely those
that have been put in evidence in the prior literature. We later went on to
exhibit the
general circumstance of a finite-time singularity within class
(\ref{exact}), which is the
situation of an incipient balloon attempting to negotiate passage through a
pinching pair of
stangation points. With arbitrarily small surface tension, the class
(\ref{exact}) flow is
unaltered until the tip of the penetrating fluid is approaching a cusp with
diverging
curvature. At this point the singular pertubation renders the stagnation
points no longer
constants of the motion. However, as soon as the pair has separated far
enough to allow the
balloon to form, the curvature is quite finite, and class (\ref{exact}) is
again correct,
save that the stagnation points are just far enough apart to allow the
minimum waisted
balloon to pass: Had we chosen initial data to have been these new
locations of stagnation
points, the $\sigma=0$ theory would have fully sufficed, and produced the
physical
solution. Following the last paragraph of Sect. 5, taking the simplest
class (\ref{exact})
solution with one pair of complex $\zeta_k$'s with Re$\alpha_k=1/2$, one
can find that
Im$\alpha_k$ with the narrowest waist, and observe the strong similarity
between the
asymptotic balloon and the best developed experimental one of \cite{58ST}.

We next observed, purely within class (\ref{exact}) however, that with
singularities
coming in clusters, each cluster well-separated in Re$\zeta$ from another,
that the solution
has asymptotic regimes, with singularities far to the left playing an
exponentially
insignificant role upon the shape of the interface, while all the others
are very close to
Re$\zeta=0$, and as we demonstated, having migrated to Im$\zeta=0$ or
$\pi$. That is, until
another cluster  of singularities at the
left arrives close to Re$\zeta=0$, at which time it joints into the
asymptotics of those
already there, the interface evolves as a single Saffman-Taylor finger. As
another cluster
arrives, that finger metamorphoses into another of a new width if
$\sum'\alpha_k$ of those
arriving differs from zero. That is, class (\ref{exact}) has the
asymptotics of always a
single finger, but of generally metamorphosing width. To establish
$\lambda=1/2$ on these
grounds requires a reason for $\sum'\alpha_k=1$ for just those
singularitieis near
Re$\zeta=0$ during the period of time of the physical experiment: with
$\sum\alpha_k=1$
for {\em all} singularities, including those arbitrarily far to the left,
demonstrates
nothing about physical pattern selection. What has fundamentally
characterized Sect. 3-5
is our focus upon temporal evolutions, accomplished via class
(\ref{exact}), by considering
this flow as a dynamical system.

Finally, we pick up on the $\pm \zeta$ symmetry, a consequence of dynamics
imposed upon
Re$\zeta=0$ of a reflection symmetric system. With any boundary fixing on
another
curve, say Re$\zeta=\xi_g(t)$, a sharp relation of $+\zeta$ to $-\zeta$
must follow,
such as with fixed pressure along a downstream line, yielding (\ref{vn3}).
This makes it
clear that arbitrarily far downstream terminations ($\xi_g\gg 1$) somehow
become entangled
with Re$\zeta\ll -1$, the domain of singularities that determine the shape
of the interface.
Although (\ref{vn3}) allows of {\em no} class (\ref{exact})$^+$ solutions,
``+" meaning
including singularities far to the right as well as those to the left of
Re$\zeta=0$, this
does not mean that theere are no solutons: this is largely the
insufficiency of finite order
class (\ref{exact}), even when extended to include higher order singular
terms. (We shall
see this in \cite{99Fei}.) However, it is clear on physical grounds that
(\ref{vn3}) is too
stringent a symmetry, and is to be replaced by myriad singularities in the
flows's analytic
continuation beyond termination. This is a serious modification of the
problem, since these
singularities are dynamical and of {\em a priori} unknown character and
locations, instead
determined by all the mechanical vagaries of the innards of a pump and so
forth, and so no
longer a physically sensibly posed problem. Instead, the physical problem
is one of boundary
geometry over just the experimentally observed body of fluid, and hence is
one of incompletely
posed geometry and data. This entails  solutions no longer unique,
requesting a physical
mechanism to select among branches etc.

Accordingly, within the machinery and formulation at hand, just one choice
lay open, which
is to consider the purely 2-D conformally well-posed problem with {\em two}
free interfaces.
The full treatment of this purely non-autonomous, non-periodic problem is
the subject of
\cite{99Fei}. We reflected some of its introductory matter into the last
paragraphs of
this paper to complete the flow of our considerations. It is worth
mentioning that there
is a class (\ref{exact}) solution with precisely on real $\zeta_k$ (and
$\alpha_k=1$),
and no others whatsoever within class (\ref{exact}).

\acknowledgments

This work has been supported in part by the Israel Science Foundation
administered
by the Israel Academy of Sciences and Humanities.

\end{document}